\documentclass[12pt]{article}

\usepackage{amsfonts,amssymb,amsmath}

\usepackage{graphicx}
\DeclareGraphicsExtensions{.pdf,.eps,.jpg}

\begin{document}

\title{\bf Selective weak measurement reveals super ergotropy}
\author{M. A. Balkanlu,
E. Faizi \thanks {E-mail:efaizi@azaruniv.ac.ir} and
B. Ahansaz
\\ {\small Physics Department, Azarbaijan Shahid Madani University, Tabriz, Iran}}\maketitle

\begin{abstract}
\noindent
The concept of ergotropy was previously introduced as the maximum extractable work from a quantum state. Its enhancement, which is induced by quantum correlation via projective measurement, was formulated as the daemonic ergotropy. In this work, we investigate the ergotropy in the presence of quantum correlation via weak measurement because of its elegant effects on the measured system. By considering a bipartite correlated quantum system consisting of main and ancillary systems, we demonstrate that the extractable work by the non-selective weak measurement on the ancilla is always equal to the situation captured by the strong measurement. However, the selective weak measurement interestingly reveals more work than the daemonic ergotropy and the ergotropy of the total system is greater than or equal to the daemonic ergotropy. Moreover, it is shown that for Bell diagonal states, at the cost of losing quantum correlation, the total extractable and thus non-local extractable works can be increased by using measurement. Also, we find that there is no direct relationship between quantum correlation and non-local extractable work for these cases.
\end{abstract}

\section{Introduction}
In quantum thermodynamics, concepts such as heat, work, thermal equilibrium, temperature, entropy and heat transfer at the atomic scale, need to be redefined and re-examined. Quantum thermodynamic concepts of heat and work were first defined by R. Alicki $\cite{Alicki}$, where the changes in the local Hamiltonian and the state of a system are associated with work and heat respectively. Recently, new definitions have been proposed as the change in the energy which is accompanied by a change in the entropy is identified as heat, while any change in the energy which does not lead to a change in the entropy is known as work $\cite{Ahmadi}$. For the case of an individual system it has been published that, similar to classical thermodynamics, the optimal extractable work is equal to its free energy change $\cite{Horodecki, Skrzypczyk}$. In general, work can only be extracted from a system whose initial state is non-passive. Thermodynamic passivity expresses that the quantum state's energy cannot be reduced by any unitary process that acts on the system. These states are so-called passive states $\cite{Sparaciari,Pusz,Perarnau}$. However, several copies of a passive state can become non-passive which allows work to be extracted from these states otherwise, the state is called completely passive $\cite{Skrzypczyk2}$. It is worth mentioning that, only the completely passive states are thermal states $\cite{Lenard}$. The maximal extractable work from a quantum non-passive state using cyclic unitary transformation is defined as ergotropy in which unitary transformation is governed by time-dependent external potentials as work sources $\cite{Allahverdian}$. There are also some ancilla-assisted protocols to work extraction processes based on quantum correlations $\cite{Francica,Perarnau2,Gonzalo,Bera}$. In a bipartite quantum system, the non-local effect on one party due to the measurement on the other is usually called quantum correlation. The non-local properties of quantum correlation make it a key element for extracting work from a global state consisting of local thermal states by using nonlocal extraction operations. Nonlocal extractable work is defined as the difference between the total and the local extractable works. G. Francica et al research $\cite{Francica}$ shows that, in a bipartite closed quantum system which is composed of the main and the ancillary systems, thanks to quantum correlation, it is possible to reach more work than the ergotropy. In this scenario, a non-interacting ancillary system is joined to the main system. Then a measurement by a set of orthogonal projectors is implemented on the ancilla which is followed by ergotropic transformation of all conditional states. This enhanced extractable work which is daemonic ergotropy is the average ergotropy of conditional states of the main system over all the possible outcomes of the measurement on the ancillary system.

The general belief that "the more information is obtained from a quantum system, the more its state is disturbed by measurement" has created a great deal of interest in finding a balance between the two $\cite{Banaszek1,Banaszek2,DAriano,Sciarrino,Andersen,Sacchi}$. Measuring the system that is necessary to observe it, will disturb the state of the system and thus its coherence. For instance, strong measurement, with complete destructing of coherence, irreversibly collapses the system state to the eigenstates of its projectors $\cite{Nielsen}$. While protecting the unknown state of a quantum system from decoherence is a key requirement for quantum information processing. Recovering the initial state after a measurement is one kind of state protection. If the interaction between the system and measuring device is weak, then the system will partially be perturbed and its coherence will not be completely annihilated. So it is possible to retrieve the initial state by undoing the effect of the measurement $\cite{Cheong}$. Such an observation of the quantum systems is known as a weak measurement $\cite{Aharonov1,Tamir,Korotkov,Jordan,Sun,Oreshkov,Dominy}$. It has recently been shown that weak measurement also helps protect quantum entanglement against decoherence based on its reversibility $\cite{Xiao,Kim}$. In a bipartite quantum system, the weak measurement performed on one of the subsystems can reveal greater quantum correlation than strong measurement $\cite{Singh}$.

The measurement on the ancillary system is the critical instrument to extract the enhanced work from the system in ancilla-assisted protocols. The reversibility of the system state under weak measurement motivated us to investigate whether applying the weak measurement would lead to a further gain in work extraction compared by projective measurement? On the hand, the relationship between geometrically defined correlation and non-local extractable work for the two-qubit pure and Bell diagonal states has been confirmed $\cite{Kevin}$. Therefore, we were interested in investigating the effect of projective measurement on total and consequently non-local extractable works and reexamining the relationship between quantum correlation and non-local extractable work in Bell diagonal states.

The rest of this paper is organized as follows. Section II briefly reviewed the definitions of ergotropy and daemonic ergotropy. In Section III, we performed non-selective weak measurement instead of projective measurement on the ancillary system and showed that the maximum extractable work is always equal to the daemonic ergotropy. But as discussed in Section IV, we demonstrated that using the selective weak measurement reveals more work than the daemonic ergotropy which we named super ergotropy. Moreover, Section V contains some geometric discussion of the local, non-local, and general extractable works for the correlated ancilla-assisted two-qubit systems. The manuscript is finished with some concluding remarks in Section VI.

\section{Ergotropy and daemonic ergotropy}

In this section, we are going to review the definitions of ergotropy and daemonic ergotropy. According to Ref. $\cite{Allahverdian}$, we assume a thermally isolated quantum system $S$ prepared in a non-passive state $\hat{\rho}(0)=\sum_{j} r_{j} |r_{j}\rangle\langle r_{j}| (r_{j}\geq r_{j+1}$  $ j=1,2,...) $ with respect to its Hamiltonian $\hat{H}=\sum_{k} \varepsilon_{k}|\varepsilon_{k}\rangle\langle \varepsilon_{k}|           (\varepsilon_{k}\leq \varepsilon_{k+1}$   $ k=1,2,...) $ which exchange work with an external macroscopic source. To extract the ergotropy, any arbitrary Hamiltonian perturbation $\hat{V}(t)$ is cyclically imposed on the system over a fixed period $[0,\tau]$. So under the action of this potential, the system undergoes a unitary transformation as $\hat{U}=\sum_{i} |\varepsilon_{i}\rangle\langle r_{i}|$ from $\hat{\rho}(0)$ to the final state $\hat{\rho}(\tau)$ with the lowest final energy. Then the system exchanges released energy as work with the external source. The ergotropy which is defined as the maximal extractable work from the state $\hat{\rho}(0)$ can be expressed as
\begin{eqnarray}
  W=Tr[\hat{\rho}(0)\Hat{H}_{s}]-Tr[\hat{\rho}(\tau)\Hat{H}_{s}]=\sum_{j,k} \varepsilon_{j}r_{k}(|\langle\varepsilon_{j}|r_{k}\rangle|^2-\delta_{jk}).
\end{eqnarray}

The ergotropy is enhanced by using the concept of quantum correlation which is obtained in Ref. $\cite{Francica}$. The procedure is that the system $S$ and a non-interacting ancillary system $A$ are prepared in a joint state\hspace{1mm}$\hat\rho_{SA}$. Then the local measurement is done on subsystem $A$ and finally, ergotropic transformation is performed for all conditional density matrices $\hat\rho_{S|A}$. The optimal extracted work from the state of the system $S$ is as follows
\begin{eqnarray}
  W_{\{\hat\Pi^A_a\}}=Tr[\hat{\rho}_{s}\Hat{H}_{s}]-\sum_{a} p_{a}Tr[\hat{U_{a}}\hat\rho_{S|a}\hat{U}^{\dagger}_{a}\hat{H}_s],
\end{eqnarray}
where $p_{a}=Tr[\hat\rho_{SA}\hat\Pi^A_a]$ is the probability of finding the system $S$ in the conditional state $ \hat\rho_{S|a}=Tr_{A}[\hat\Pi^A_a\hat\rho_{SA}\hat\Pi^A_a]/p_{a}$, $\hat{U_{a}}$ is the cyclic unitary process that transforms $\hat\rho_{S|a}$ in order to minimize its internal energy and ${\{\hat\Pi^A_a\}}$ is a set of orthogonal projectors of measurement. The extracted work $W_{\{\hat\Pi^A_a\}}$ which is known as Daemonic Ergotropy is calculated as
\begin{eqnarray}
  W_{\{\hat\Pi^A_a\}}=Tr[\hat{\rho}_{s}\Hat{H}_{s}]-\sum_{a} p_{a}\sum_{k}r^{a}_{k}\varepsilon_{j},
\end{eqnarray}
where $r^{a}_{k}$ are the eigenvalues of $\hat\rho_{S|a}$. The state $\hat\rho_{S}$ can be written as $\hat\rho_{S}=\sum_{a}p_{a}\hat\rho_{S|a}=\sum_{a}p_{a}\sum_{k}r^{a}_{k}|r^{a}_{k}\rangle\langle r^{a}_{k}|$ which implies $r_{k}=\sum_{a}p_{a}\sum_{j}r^{a}_{j}|\langle r_{k}|r^{a}_{j}\rangle|^2$and finally the difference $W_{\{\hat\Pi^A_a\}}-W$ reads as
\begin{eqnarray}
  W_{\{\hat\Pi^A_a\}}-W=\sum_{k}\varepsilon_{k}(r_{k}-\sum_{a}p_{a}r^{a}_{k})=\sum_{a}p_{a}\sum_{k,j}r^{a}_{j}\varepsilon_{k}(|\langle r_{k}|r^{a}_{j}\rangle|^2-\delta_{jk}).
\end{eqnarray}
This is the average of all the conditional states ergotropy relative to Hamiltonian $\hat{H}=\sum_{k} \varepsilon_{k}|r_{k}\rangle\langle r_{k}|$ which are non-negative by definition. From the positivity of Eq. (4), we have $W_{\{\hat\Pi^A_a\}}\geq W$. When the subsystems are initially uncorrelated as  $\hat\rho_{SA}=\hat\rho_{S}\otimes\hat\rho_{A}$, we have $\hat\rho_{S|a}=\hat\rho_{S}$ for any set $\{\hat\Pi^A_a\}$ and outcome $a$, which means there would be no gain in work extraction and  $W_{\{\hat\Pi^A_a\}}=W$.

\section{Non-selective weak measurement and daemonic ergotropy}
The notion of weak measurement with post-selection was first introduced in Ref. $\cite{Aharonov1}$, which leads to the definition of weak values of the measured observable. Weak values can be calculated by using the pre and post-selected quantum systems and can be beyond the permissible values for observable. We will not be considering the measurements of this type. Throughout this paper, we will use another expression for weak measurement in terms of measurement operators which is given by Oreshkov and Brun for the weak measurement in Ref. $\cite{Oreshkov}$. In this section, we have shown that by applying the mentioned restrictions on the system dimension and projectors, the same results as in the previous section are obtained by non-selective weak measurement on the ancillary system in a joint state $\hat\rho_{SA}$. Nevertheless, its advantage is to maintain the correlation between subsystems.
Following the second formalism of the weak measurement, the operators of weak measurement have the form as
\begin{align}
  \hat {P}(x)=\sqrt{b_{0}(x)} \hat\Pi_0+\sqrt{b_{1}(x)} \hat\Pi_1,
\end{align}
\begin{align}
  \hat {P}(-x)=\sqrt{b_{0}(-x)} \hat\Pi_0+\sqrt{b_{1}(-x)} \hat\Pi_1,
\end{align}
and the POVM elements $\hat {E(x)}$, $\hat {E(-x)}$ for this measurement is given as
\begin{eqnarray}
  \hat {E}(x)=\hat {p}^\dagger (x)\hat {p}(x)=b_{0}(x)\hat\Pi_{0}+b_{1}(x)\hat\Pi_{1},
\end{eqnarray}
\begin{eqnarray}
  \hat {E}(-x)=\hat {p}^\dagger (-x)\hat {p}(-x)=b_{0}(-x)\hat\Pi_{0}+b_{1}(-x)\hat\Pi_{1},
\end{eqnarray}
where $b_{0}(x)=(1-\mathrm{tanh} x)/2, b_{1}(x)=(1+\mathrm{tanh} x)/2 $ and $x\in [0,\infty]$ is the parameter to control the strength of measurement.
$\hat\Pi_{0}$ and $\hat\Pi_{1}$ are two orthogonal projectors with $\hat\Pi_{0}+\hat\Pi_{1}=I$. Note that in the limit of $x\rightarrow 0$, the operators $\hat {P}(x)$ and $\hat {P}(-x)$ are proportional to the identity which means no measurement is done. In the limit of strong measurement, we have $\mathrm{lim}_{x\rightarrow \infty} \hat {P}(-x)=\hat\Pi_{0}$ and $\mathrm{lim}_{x\rightarrow \infty} \hat {P}(x)=\hat\Pi_{1}$. These operators also satisfy
$\hat {P}(x)\hat {P}(y)\propto \hat {P}(x+y)$ and $[\hat {P}(x),\hat {P}(y)]=I$.
We can write the extractable work from the state of systems average ergotropies of two possible outcomes of measurement
\begin{eqnarray}
  W_{\{\hat P^A_{\pm x}\}}=Tr[\hat{\rho}_{s}\Hat{H}_{s}]-\sum_{y=x, -x}p_{y}Tr[\hat{U_{y}}\hat\rho_{S|y}\hat{U}^{\dagger}_{y}\hat{H}_s].
\end{eqnarray}

$\bf{Theorem \ 1.}$ For any system $S$ and ancillary system $A$ prepared in a state $\hat\rho_{SA}$, by performing weak measurement on the ancillary system, we always have $W_{\{\hat P^A_{\pm x}\}}=W_{\{\hat \Pi^A_{a}\}}$.\\

$\bf{Proof. \ }$ We can write probabilities of measurement as below
\begin{align}
p_{x}&=Tr[(\hat I\otimes \hat {E}(x) )\hat\rho_{SA}] \\    \nonumber
&=Tr\{[\hat I\otimes (b_{0}(x)\hat\Pi_{0}+b_{1}(x)\hat\Pi_{1})]\hat\rho_{SA}\}=b_{0}(x)p_{0}+b_{1}(x)p_{1},
\end{align}
and $\hat\rho_{S|x}$ which is related to $\hat\rho_{S|a}$ is given by
\begin{align}
\hat\rho_{S|x}=\frac{Tr_{A}[(\hat I\otimes \hat {E}(x) )\hat\rho_{SA}]}{p(x)}=\frac{\sum_{a}b_{a}(x)p_{a}\hat\rho_{S|a}}{p(x)}.
\end{align}
By considering equations (10) and (11), Eq. (9) reduces as follows
\begin{eqnarray}
W_{\{\hat P^A_{\pm x}\}}=Tr[\hat{\rho}_{s}\Hat{H}_{s}]-\sum_{a}\sum_{y=x, -x} b_{a}(y) p_{a}Tr[\hat{U_{a}}\hat\rho_{S|a}\hat{U}^{\dagger}_{a}\hat{H}_s].
\end{eqnarray}
Similar to Eq. (4), the difference $W_{\{\hat P^A_{\pm x}\}}-W$ is given by
\begin{eqnarray}
 W_{\{\hat P^A_{\pm x}\}}-W=\sum_{k}\varepsilon_{k}(r_{k}-\sum_{a}\sum_{y=x, -x}b_{a}(y) p_{a} r^{a}_{k}).
\end{eqnarray}
By using the fact $\sum_{y=x, -x}b_{a}(y)=1$, we have $W_{\{\hat P^A_{\pm x}\}}=W_{\{\hat \Pi^A_{a=0, 1}\}}$. Which shows the equality of maximum extractable works for the two sets of measurement operators.

\section{Selective weak measurement and super ergotropy}
The daemonic ergotropy can be obtained via weak measurement on the ancillary system to avoid quantum correlation loss. This fact motivated the authors to optimize the extractable work which led to the investigation of selective weak measurement i.e. by applying one of its operators. We proceed with the work extraction by measuring the ancillary system with one of the weak measurement operators. So the conditional density matrix of the system $S$ given the measurement operator $\hat P{(\pm x)}$ on the system $A$ can be written as
\begin{eqnarray}
\hat\rho_{S|\pm x}=\frac{\sum_{a}b_{a}(\pm x)p_{a}\hat\rho_{S|a}}{p_{(\pm x)}}
\end{eqnarray}
where
\begin{eqnarray}
p_{\pm x}  =b_{0}(\pm x)p_{0}+b_{1}(\pm x)p_{1}.
\end{eqnarray}
The extractable work from the state of system $S$ reads as
\begin{eqnarray}
W_{\hat P^A_{\pm x}}=Tr[\hat{\rho}_{s}\Hat{H}_{s}]-\frac{1}{p_{(\pm x)}}\sum_{a=0}^{1} b_{a}(\pm x) p_{a}Tr[\hat{U_{a}}\hat\rho_{S|a}\hat{U}^{\dagger}_{a}\hat{H}_s].
\end{eqnarray}
So the difference between the maximal extractable work by using non-selective (projective measurement) and selective (weak measurement) on the subsystem $A$ can be obtained as
\begin{align}
\delta_{W(\pm x)} =W_{\hat P^A_{\pm x}}-W_{\{\hat \Pi^A_{a}\}}&=\sum_{a=0}^{1}(1-\frac{b_{a}(\pm x)}{p_{(\pm x)}})Tr[\hat{U_{a}}\hat\rho_{S|a}\hat{U}^{\dagger}_{a}\hat{H}_s]\\  \nonumber
&=\pm C_{\pm}(Tr[\hat{U_{0}}\hat\rho_{S|0}\hat{U}^{\dagger}_{0}\hat{H}_s]-Tr[\hat{U_{1}}\hat\rho_{S|1}\hat{U}^{\dagger}_{1}\hat{H}_s]),
\end{align}
where $C_{\pm}=\frac{2p_{0}p_{1}\mathrm{tanh} x}{1\pm(p_{1}-p_{0})\mathrm{tanh} x}$ and signs $+$ and $-$ are related to $\hat{P}(+x)$ and $\hat{P}(-x)$ respectively. These coefficients are positive and depend on the probability of measurement outcomes and its intensity.
\\

$\bf{Theorem \ 2.}$ For any system $S$ and ancillary system $A$ prepared in a state $\hat\rho_{SA}$, by performing the selective weak measurement on the ancillary system we always have $W_{\hat P^A_{\pm x}}\ge W_{\{\hat \Pi^A_{a=0, 1}\}}$.\\

$\bf{Proof. \ }$ Considering Eq. (17), we select measurement operator $P(x)$ when $Tr[\hat{U_{0}}\hat\rho_{S|0}\hat{U}^{\dagger}_{0}\hat{H}_s]$ is greater than $Tr[\hat{U_{1}}\hat\rho_{S|1}\hat{U}^{\dagger}_{1}\hat{H}_s]$ and we chose measurement operator $P(-x)$ when $Tr[\hat{U_{1}}\hat\rho_{S|1}\hat{U}^{\dagger}_{1}\hat{H}_s]$ is greater than $Tr[\hat{U_{0}}\hat\rho_{S|0}\hat{U}^{\dagger}_{0}\hat{H}_s]$ in order to $\delta_{W(\pm x)}$ be positive. If two traces are equal, the extractable work by both measurements is the same.

Therefore, depending on the final energy of conditional states $\hat\rho_{S|a}$, extractable work greater than or equal to the demonic ergotropy can be achieved. We called extractable work $W_{\hat P^A_{\pm x}}$ as super ergotropy.

\subsection{Example}
In the following, we introduce some examples to make the theorem more clear.
If we consider the pure states $\hat{\rho}_1=|\psi\rangle_{SA}\langle\psi|$ with $|\psi\rangle_{SA}=\mathrm{cos}(\theta)|00\rangle+\mathrm{sin}(\theta)|00\rangle$ and $\hat{\rho}_2=|\phi\rangle_{SA}\langle\phi|$ with $|\phi\rangle_{SA}=\frac{1}{\sqrt3}(|11\rangle+|10\rangle+|01\rangle)$ the maximum work which can be extracted by using the selective weak measurement is the same as with the demonic ergotropy. Here we will show the ability of the selective weak measurement to gain more work than demonic ergotropy. As another example, we consider the mixed state $\hat\rho_{SA}=\frac{1}{3}(|11\rangle\langle11|+|10\rangle\langle10|+|01\rangle\langle01|)$ which its daemonic ergotropy is calculated from Eq. (2) as $W_{\{\hat \Pi^A_{a=0, 1}\}}=\frac{\varepsilon_{1}-\varepsilon_{0} }{3}=\frac{\Delta \varepsilon }{3}$, where $\varepsilon_{1}$ and $\varepsilon_{0}$ are the energy of levels $1$ and  $0$ of the system $S$, respectively. By measuring the ancillary system via operator $P(x)$ followed by the ergotropic transformation, the maximal extractable work reads as
 \begin{eqnarray}
W_{\hat P^A_{+x}}=\frac{3+\mathrm{tanh} x}{3-\mathrm{tanh} x}(\Delta \varepsilon /3).
\end{eqnarray}
The Fig. 1 shows $W_{\hat P^A_{+x}}/W_{\{\hat \Pi^A_{a}\}}$ as a function of measurement strength $x$ which is greater than one. The more strongly one perturbs the subsystem $A$, the more work can be revealed from system $S$. So for sufficiently large values of $x$, we have $W_{\hat P^A_{+x}}=2W_{\{\hat \Pi^A_{a}\}}$. In this case, selective strong measurement with projector $\Pi^A_1$ emerges instead of  $P(+x)$. However, when we set $x=0$, the extractable works $W_{\hat P^A_{+x}}$ and $W_{\{\hat \Pi^A_{a}\}}$ became equal.

\section{Measurement effect on total and non-local extractable works}
Non-local extractable work of a joint system is defined as the difference between the total and the local extractable works $\cite{Kevin}$. We are also interested in measuring the total system by a set of projective operators $\{\hat {\Pi}_{ij}\}$. We define a classical non-local extractable work as the difference between the average ergotropy of projective measurement on the general system and the local extractable work as
\begin{eqnarray}
W^{\{\hat {\Pi}_{ij}\}}_{Non-loc}=W^{\{\hat {\Pi}_{ij}\}}_{tot}-W_{loc}.
\end{eqnarray}
The definition above enables us to compare the extractable work from quantum and classically correlated states and examine the role of measurement in this scenario.

In this section, we consider a general two-qubit system with computational bases $\{|11\rangle, |10\rangle, \\|01\rangle, |00\rangle\}$ which consists of non-interacting main and ancillary subsystems. The density matrix of such a system can be written in geometrically Bloch representation as
\begin{eqnarray}
\hat{\rho}_{SA}=\frac{1}{4}(\hat{I}+\sum_{i=1}^{3} s_{i}\hat{\sigma}_{i}\otimes \hat{I}+\sum_{j=1}^{3} r_{j}\hat{I}\otimes \hat{\sigma}_{j}+\sum_{i,j=1}^{3}t_{ij}\hat{\sigma}_{i}\otimes \hat{\sigma}_{j}),
\end{eqnarray}
where $\hat{I}$ is  the identity matrix, $s_{i}=Tr[\hat{\rho}_{SA}(\hat{\sigma}_{i}\otimes \hat{I})]$, $r_{j}=Tr[\hat{\rho}_{SA}(\hat{I}\otimes \hat{\sigma}_{j})]$ and $t_{ij}=Tr[\hat{\rho}_{SA}(\hat{\sigma}_{i}\otimes \hat{\sigma}_{j})]$  is the $3\times 3$ correlation matrix and $\hat{\sigma}_{i}(\hat{\sigma}_{j})$ are the Pauli matrices.
Now, we perform measurement on the ancillary system with the set of orthogonal projectors $\hat{\Pi}_{1}$ and $\hat{\Pi}_{0}$ respectively. The conditional density matrix of the system $S$ given projector $\hat{\Pi}_{1}$ on the system $A$ reads as
\begin{eqnarray}
  \hat{\rho}_{{S|1}}=\frac{1}{4}[(1+r_{3})\hat{I}+\sum_{i}(s_{i}+t_{i3})\hat{\sigma}_{i}]=\lambda_{+}|\lambda_{+}\rangle\langle \lambda_{+}|          +\lambda_{-}|\lambda_{-}\rangle\langle \lambda_{-}|,
\end{eqnarray}
with
\begin{eqnarray}
\lambda_{\pm}=\frac{1+r_{3}\pm|\vec s+\vec t_{i3}|}{4},
\end{eqnarray}
where $|\vec s+\vec t_{i3}|=\sqrt{(s_{1}+t_{13})^2+(s_{2}+t_{23})^2+(s_{3}+t_{33})^2}$. The conditional density matrix of the system $S$ given projector $\hat{\Pi}_{0}$ on the system $A$ reads as
\begin{eqnarray}
  \hat{\rho}_{{S|0}}=\frac{1}{4}[(1-r_{3})\hat{I}+\sum_{i}(s_{i}-t_{i3})\hat{\sigma}_{i}]=\lambda_{+}^\prime|\lambda_{+}^\prime\rangle\langle \lambda_{+}^\prime|+\lambda_{-}^\prime|\lambda_{-}^\prime\rangle\langle \lambda_{-}^\prime|,
\end{eqnarray}
with
\begin{eqnarray}
\lambda_{\pm}^\prime=\frac{1+r_{3}\pm|\vec s-\vec t_{i3}|}{4},
\end{eqnarray}
where $|\vec s-\vec t_{i3}|=\sqrt{(s_{1}-t_{13})^2+(s_{2}-t_{23})^2+(s_{3}-t_{33})^2}$. The unitary operators of  the ergotropic transformation are $\hat{U}_{1}=|1\rangle\langle \lambda_{+}|+|0\rangle\langle \lambda_{-}|$ and $\hat{U}_{0}=|1\rangle\langle \lambda_{+}^\prime|+|0\rangle\langle \lambda_{-}^\prime|$ which gives the daemonic ergotropy as
\begin{eqnarray}
W_{\{\hat \Pi^A_{a}\}}=(2s_{3}+|\vec s+\vec t_{i3}|+|\vec s-\vec t_{i3}|)\frac{\Delta\varepsilon}{4}.
\end{eqnarray}
On the other hand, the reduced density matrix of the system $S$ is given by $\hat\rho_{S}=Tr_{A}(\hat\rho_{SA})=\frac{1}{2}(\hat{I}+\sum_{i}s_{i}\hat{\sigma}_{i})$. Due to the zero Hamiltonian of the ancillary system, the local extractable work will be only the ergotropy of the main system as $W_{loc}=(s_{3}+s)\frac {\Delta\varepsilon}{2}$ $\cite{Kevin}$.
As mentioned in section II, we have $W_{\{\hat \Pi^A_{a}\}}\ge W_{loc}$, which is compatible with this geometric interpretation which states for two arbitrary vectors $A$ and $B$, $|\vec A+\vec B|+|\vec A-\vec B|\ge 2|\vec A|$. Classical non-local extractable work from Eq. (19) obtains as $W^{\{\hat {\Pi}_{ij}\}}_{Non-loc}=(1-s)\frac {\Delta\varepsilon}{2}$  (Appendix A). As an interesting result,  when the pre-measurement state of the main system is pure,  $W^{\{\hat {\Pi}_{ij}\}}_{Non-loc}$ is always equal to zero, but for the mixed pre-measurement state of this system, it will be greater than or equal to zero.

Up to local unitary equivalence, a general two-qubit state is always reducible to a state at the Bloch normal form as \cite{Shunlong}
\begin{eqnarray}
\hat{\rho}_{SA}=\frac{1}{4}(\hat{I}+\vec s\hat{\sigma}\otimes \hat{I}+\hat{I}\otimes \vec r \hat{\sigma}_{j}+\sum_{i=1}^{3}c_{i}\hat{\sigma}_{i}\otimes \hat{\sigma}_{i}),
\end{eqnarray}
where $\vec s$ and $\vec r$ are the Bloch vectors of the reduced density matrices of the systems $S$ and $A$ respectively and $c_{i}$ are the real numbers such that $|c_{i}|\le 1$. We assume that $\vec s=\vec r=0$ to reduce it to the Bell-diagonal states
\begin{align}
\hat{\rho}_{SA}&=\frac{1}{4}(\hat{I}+\sum_{i=1}^{3}c_{i}\hat{\sigma}_{i}\otimes \hat{\sigma}_{i})=\sum_{i=1}^{3}\lambda_{i}|\lambda_{i}\rangle\langle \lambda_{i}|,
\end{align}
with $\lambda_{0}=1-c_{1}-c_{2}-c_{3}, \lambda_{1}=1-c_{1}+c_{2}+c_{3}, \lambda_{2}=1
+c_{1}-c_{2}+c_{3}$ and $\lambda_{3}=1+c_{1}+c_{2}-c_{3}$.
The reduced density matrices are maximally mixed states and cannot be transformed by unitary operations, so the local extractable work of $\hat{\rho}_{SA}$ will be zero $W_{loc}=0$. The conditional density matrices of the system  given the projective measurement operators $\hat{\Pi}_{i} (i=0, 1)$ on the ancillary system can be obtained as
\begin{eqnarray}
  \hat{\rho}_{S|1}=\frac{1}{2}[\hat{I}+c_{3}\hat{\sigma}_{3}]=\frac{1}{2}[(1+c_{3})|1\rangle\langle1|          +(1-c_{3})|0\rangle\langle0|]  \text{ with} \hspace{1mm}p_{1}=\frac{1}{2},
\end{eqnarray}
\begin{eqnarray}
  \hat{\rho}_{S|0}=\frac{1}{2}[\hat{I}-c_{3}\hat{\sigma}_{3}]=\frac{1}{2}[(1-c_{3})|1\rangle\langle1|          +(1+c_{3})|0\rangle\langle0|] \text{ with} \hspace{1mm}    p_{0}=\frac{1}{2}.
\end{eqnarray}
So the daemonic ergotropy of the system $S$ is given by
\begin{eqnarray}
    W_{\{\hat \Pi^A_{a}\}}=|c_{3}|\frac {\Delta\varepsilon}{2}.
\end{eqnarray}
On the other hand, the total extractable work is obtained by the ergotropic transformation $\hat{U}_{SA}$ on the state $\hat{\rho}_{SA}$. Depending on which $|c_{i}|$ is greatest than the others, the corresponding cyclic unitary transformation is required to extract work from the joint system state. The ergotropy of the total system is given by

\begin{eqnarray}
 W_{tot}= \begin {cases}

|c_{1}|\Delta\varepsilon/2 & \text{ with} \hspace{3mm} \hat{U}_{1}=|11\rangle\langle\lambda_{0}|+|10\rangle\langle\lambda_{1}|+|01\rangle\langle\lambda_{2}|+|00\rangle\langle\lambda_{3}|,\\ |c_{2}|\Delta\varepsilon/2 & \text{ with} \hspace{3mm} \hat{U}_{2}=|11\rangle\langle\lambda_{0}|+|10\rangle\langle\lambda_{2}|+|01\rangle\langle\lambda_{1}|+|00\rangle\langle\lambda_{3}|,\\|c_{3}|\Delta\varepsilon/2 & \text{ with} \hspace{3mm} \hat{U}_{3}=|11\rangle\langle\lambda_{0}|+|10\rangle\langle\lambda_{3}|+|01\rangle\langle\lambda_{1}|+|00\rangle\langle\lambda_{2}|,

\end {cases}
\end{eqnarray}
which is greater than or equal to the daemonic ergotropy $W_{\{\hat \Pi^A_{a}\}}$. Now, we perform the projective measurement on the total system by a set of $\hat\Pi_{ij}$ and calculate the conditional density matrices as

\begin{eqnarray}
\begin{array}{c}
\hat{\rho}^{SA}_{\hat {\Pi}_{11}}=\frac{Tr(\hat{\Pi}_{11}\hat{\rho}_{SA})}{p_{11}}=|11\rangle\langle11|\hspace{3mm} \text{ with} \hspace{3mm} p_{11}=\frac {1}{4}(1+c_{3}),\\
\hat{\rho}^{SA}_{\hat {\Pi}_{10}}=\frac{Tr(\hat{\Pi}_{10}\hat{\rho}_{SA})}{p_{10}}=|10\rangle\langle10|\hspace{3mm}\text{ with} \hspace{3mm} p_{10}=\frac {1}{4}(1-c_{3}),\\
\hat{\rho}^{SA}_{\hat {\Pi}_{01}}=\frac{Tr(\hat{\Pi}_{01}\hat{\rho}_{SA})}{p_{01}}=|01\rangle\langle01|\hspace{3mm}\text{ with} \hspace{3mm} p_{01}=\frac {1}{4}(1-c_{3}),\\
\hat{\rho}^{SA}_{\hat {\Pi}_{00}}=\frac{Tr(\hat{\Pi}_{00}\hat{\rho}_{SA})}{p_{00}}=|00\rangle\langle00|\hspace{3mm}\text{ with} \hspace{3mm} p_{00}=\frac {1}{4}(1+c_{3}).\\
\end{array}
\end{eqnarray}
The energy of each conditional state is reduced to $\varepsilon_{0}$  by using the cyclic unitary operators, so the average ergotropy of all the conditional states reads as follows
\begin{eqnarray}
W^{\{\hat{\Pi}_{ij}\}}_{tot}=Tr[\hat{\rho}_{SA}\hat{H}_{SA}]-\sum_{i,j} p_{ij}Tr[\hat{U_{ij}}\hat\rho_{SA|ij}\hat{U}^{\dagger}_{ij}\hat{H}_{SA}]=\Delta\varepsilon/2,
\end{eqnarray}
which is greater than or equal to the total extractable work of the joint system state. It means we can extract more work by performing the projective measurement on the state $\hat\rho_{SA}$ than both the total extractable work and the daemonic ergotropy. On the other hand, since $W_{loc}=0$ therefore $W^{\{\hat{\Pi}_{ij}\}}_{Non-loc}=W^{\{\hat{\Pi}_{ij}\}}_{tot}$ . Thus for the post-measurement state $\hat{\rho}_{SA}=\sum_{i,j} p_{ij}\hat\rho_{SA|ij}$, the non-local extractable work is maximized, however, the quantum correlation is completely lost. Before such measurement, by using a measure which is introduced in Ref. $\cite{Hui}$, one can get the quantum correlation as
\begin{eqnarray}
    C(\hat{\rho})=\sqrt{2-\sqrt{4-[c^2_{1}c^2_{2}+c^2_{1}c^2_{3}+c^2_{2}c^2_{3}}}].
\end{eqnarray}
As we see from Eq. (34), the quantum correlation depends on all $c_{i}$s $(i=1, 2, 3)$, however non-local extractable work $W_{non-loc}$ is only proportional to one of them. In general, there is not an explicit relationship between these two quantities. The Fig. 2 shows the quantum correlation $C(\hat{\rho})$ as the function of $c_{1}/c_{2}$ and $c_{3}/c_{2}$. As one can see this quantity is maximal when all $c_{i}$s tend to $1$.

\subsection{Example}
As an example of Bell-diagonal states, we ploted $W^{\{\hat{\Pi}_{ij}\}}_{Non-loc}$ and $W_{Non-loc}$ in Fig. 3 for $c_{1}=\frac{1}{2}$, $c_{2}=-\frac{1}{2}$ and $c_{3}=\mathrm{sin}(\theta)$. For post measurement classical correlated state $\hat{\rho}_{SA}=\frac{1}{4}\sum_{j,k}(1+(-1)^{j+k}c_{3})|jk\rangle\langle jk| (j,k=0,1)$, the Fig. 3 shows that $W^{\{\hat{\Pi}_{ij}\}}_{Non-loc}$ is greater than  $W_{Non-loc}$ except in $\theta =\frac{\pi}{2}$ which are equal. In Fig. 4, we ploted the quantum correlation in terms of $\theta$. It shows the correlation of the initial state before the work extraction process which is calculated as $\frac{1}{2}\sqrt{8-\sqrt{63-8sin^2(\theta)}}$. As one can see the quantum correlation is maximized at $\theta =\frac{\pi}{2}$, which specifies the extremum point for two types of non-local extractable work in Fig. 3.

\section{Conclusions}
In this paper, we studied the concept of gain in work extraction in closed quantum systems in the presence of quantum correlation through the performance of the weak measurement. We considered a bipartite quantum system and showed that the extractable work induced by the weak measurement on the ancillary subsystem was always equal to the daemonic ergotropy captured by the projective measurement independently of the strength of the weak measurement. As an interesting result, we extracted more work by using the selective weak measurements on the ancillary system which we called super ergotropy. It depends on the final internal energy of the conditional principal system states to specify which of the weak measurement operators to be imposed. We also provided a geometric confirmation for daemonic ergotropy in the case of general two-qubit states. In this regard, we defined a non-local extractable work for bipartite systems which are classically correlated and showed that if the state of the main subsystem is pure, regardless of whether the general two-qubit state after the perfect projective measurement on the total system has a classical correlation or not, the defined classical non-local work is equal to zero, and if its state is mixed, this work can be non-zero. Moreover, we investigated total system ergotropy in the case of Bell diagonal states and showed that in the effect of perfect projective measurement on the total system, the average ergotropy of conditional states was greater than total extractable work in the cost of losing quantum correlation between subsystems. It can be generalized daemonic ergotropy. Our research showed no explicit relationship between non-local extractable work and the quantum correlation of subsystems in the pre-measurement state of the last-mentioned total system.

\section*{Data availability}
The present study's data are available from the corresponding author upon a reasonable request.

\section*{Acknowledgments}
The authors would like to thank Abbas Ektesabi for very useful comments and advice.

\section*{Appendix A}
The Hamiltonian of the two-qubit state is
\begin{eqnarray}
\hat{H}=\hat{H}_{A}\otimes\hat{I}.
\end{eqnarray}
So initial energy of the total system is given by
\begin{align}
Tr[\hat{\rho}_{SA}\hat{H}_{SA}]&=\frac{1}{4}Tr[(\hat{I}+\sum_{i=1}^{3} s_{i}\hat{\sigma}_{i}\otimes \hat{I}+\sum_{j=1}^{3} r_{j}\hat{I}\otimes \hat{\sigma}_{j}+\sum_{i,j=1}^{3}t_{ij}\hat{\sigma}_{i}\otimes \hat{\sigma}_{j})(\hat{H}_{A}\otimes\hat{I})]\\ \nonumber
&=\frac{1}{2}(\varepsilon_{0}+\varepsilon_{1}+s_{3}\Delta\varepsilon).
\end{align}
By measuring the two-qubit system via a set of projective operators $\hat{\Pi}_{ij}(i,j=0, 1)$ and calculating the average ergotropy overall outcomes of measurement can be obtained as
\begin{align}
\begin{array}{c}
\hat{\rho}^{SA}_{\hat {\Pi}_{11}}=\frac{Tr(\hat{\Pi}_{11}\hat{\rho}_{SA})}{p_{11}}=|11\rangle\langle11| \hspace{3mm}\text{ with} \hspace{3mm} p_{11}=\frac {1}{4}(1+s_{3}+r_{3}+t_{33}),\\
\hat{\rho}^{SA}_{\hat {\Pi}_{10}}=\frac{Tr(\hat{\Pi}_{10}\hat{\rho}_{SA})}{p_{10}}=|10\rangle\langle10|\hspace{3mm}\text{ with} \hspace{3mm} p_{10}=\frac {1}{4}(1+s_{3}-r_{3}-t_{33}),\\
\hat{\rho}^{SA}_{\hat {\Pi}_{01}}=\frac{Tr(\hat{\Pi}_{01}\hat{\rho}_{SA})}{p_{01}}=|01\rangle\langle01|\hspace{3mm}\text{ with} \hspace{3mm} p_{01}=\frac {1}{4}(1-s_{3}+r_{3}-t_{33}),\\
\hat{\rho}^{SA}_{\hat {\Pi}_{00}}=\frac{Tr(\hat{\Pi}_{00}\hat{\rho}_{SA})}{p_{00}}=|00\rangle\langle00|\hspace{3mm}\text{ with} \hspace{3mm} p_{00}=\frac {1}{4}(1-s_{3}-r_{3}+t_{33}).\\
\end{array}
\end{align}
The cyclic unitary operators which minimize the energy of the conditional state are given by
\begin{eqnarray}
\begin{array}{c}
\hat{U}_{11}=|01\rangle\langle11|+|11\rangle\langle01| \text {or}\hspace{1mm} \hat{U}_{11}=|00\rangle\langle11|+|11\rangle\langle00|,\\
\hat{U}_{10}=|01\rangle\langle10|+|10\rangle\langle01| \hspace{1mm}\text {or} \hspace{1mm} \hat{U}_{10}=|00\rangle\langle10|+|10\rangle\langle00|,\\
\hat{U}_{01}=\hat{I},\\
\hat{U}_{00}=\hat{I}.
\end{array}
\end{eqnarray}
All the cyclic unitary operators lead the energy of the total system to the lowest eigenstate of the Hamiltonian. So the total extractable work is
\begin{align}
W^{\{\hat{\Pi}_{ij}\}}_{tot}&=Tr[\hat{\rho}_{SA}\hat{H}_{SA}]-\sum_{i,j} p_{i,j}Tr[\hat{U_{ij}}\hat\rho_{SA|ij}\hat{U}^{\dagger}_{ij}\hat{H}_{SA}]\\ \nonumber
&=(1+s_{3})\Delta\varepsilon/2.
\end{align}
So the classical non-local extractable work abstains as

\begin{eqnarray}
W^{\{\hat{\Pi}_{ij}\}}_{Non-loc}=W^{\{\hat{\Pi}_{ij}\}}_{tot}-W=(1-s)\Delta\varepsilon/2.
\end{eqnarray}

\newpage

\newpage
Fig. 1. $\frac {W_{\hat P^A_{+x}}}{W_{\{\hat \Pi^A_{a}\}}}$ as a function of measurement strength $x$.
\begin{figure}
\centering
\includegraphics[width=300 pt]{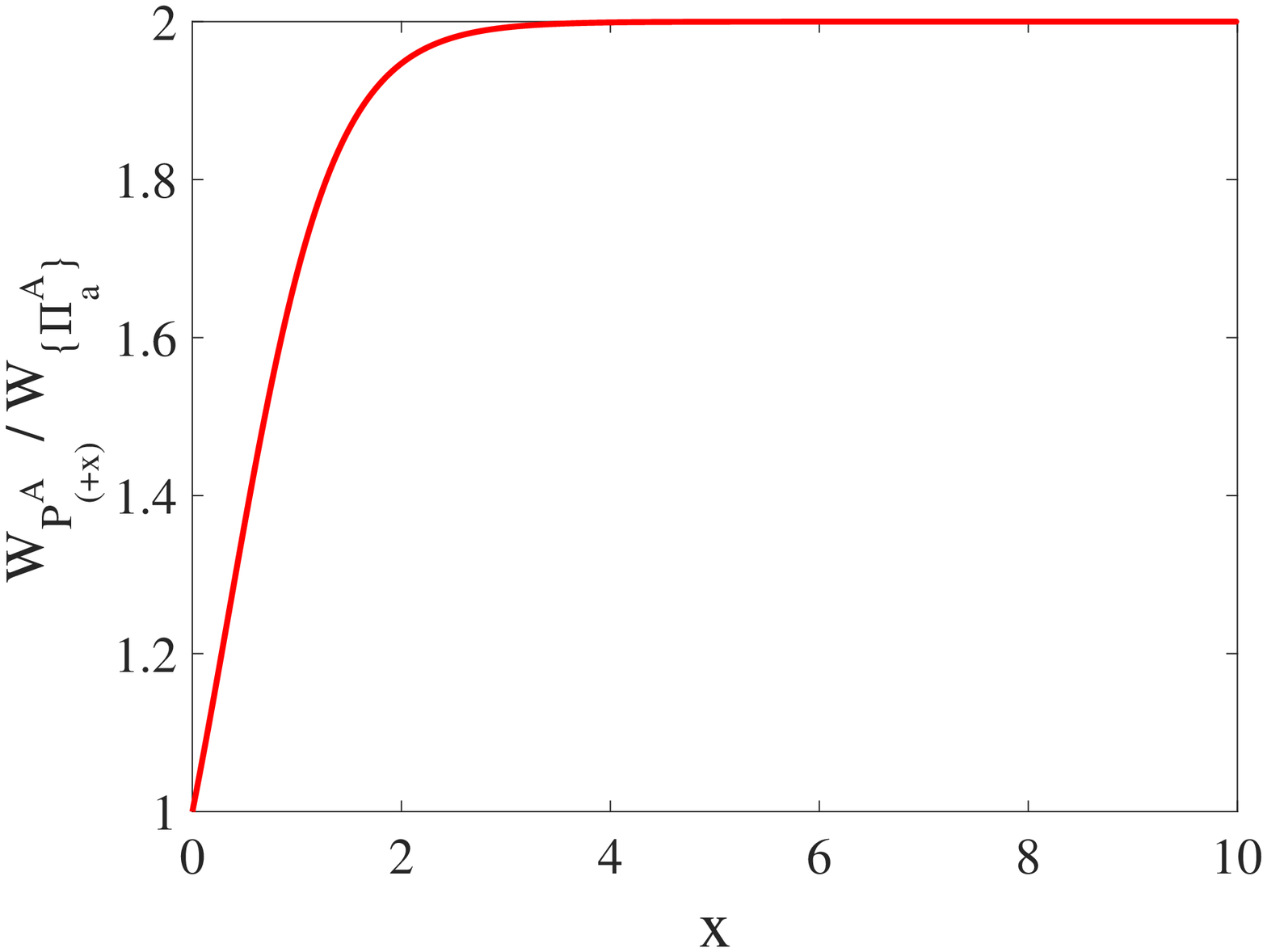}
\caption{} \label{Fig1a}
\end{figure}

\newpage
Fig. 2. The quantum correlation of Bell-diagonal state as the function of $c_{1}/c_{2}$ and $c_{3}/c_{2}$.
\begin{figure}
\centering
\includegraphics[width=300 pt]{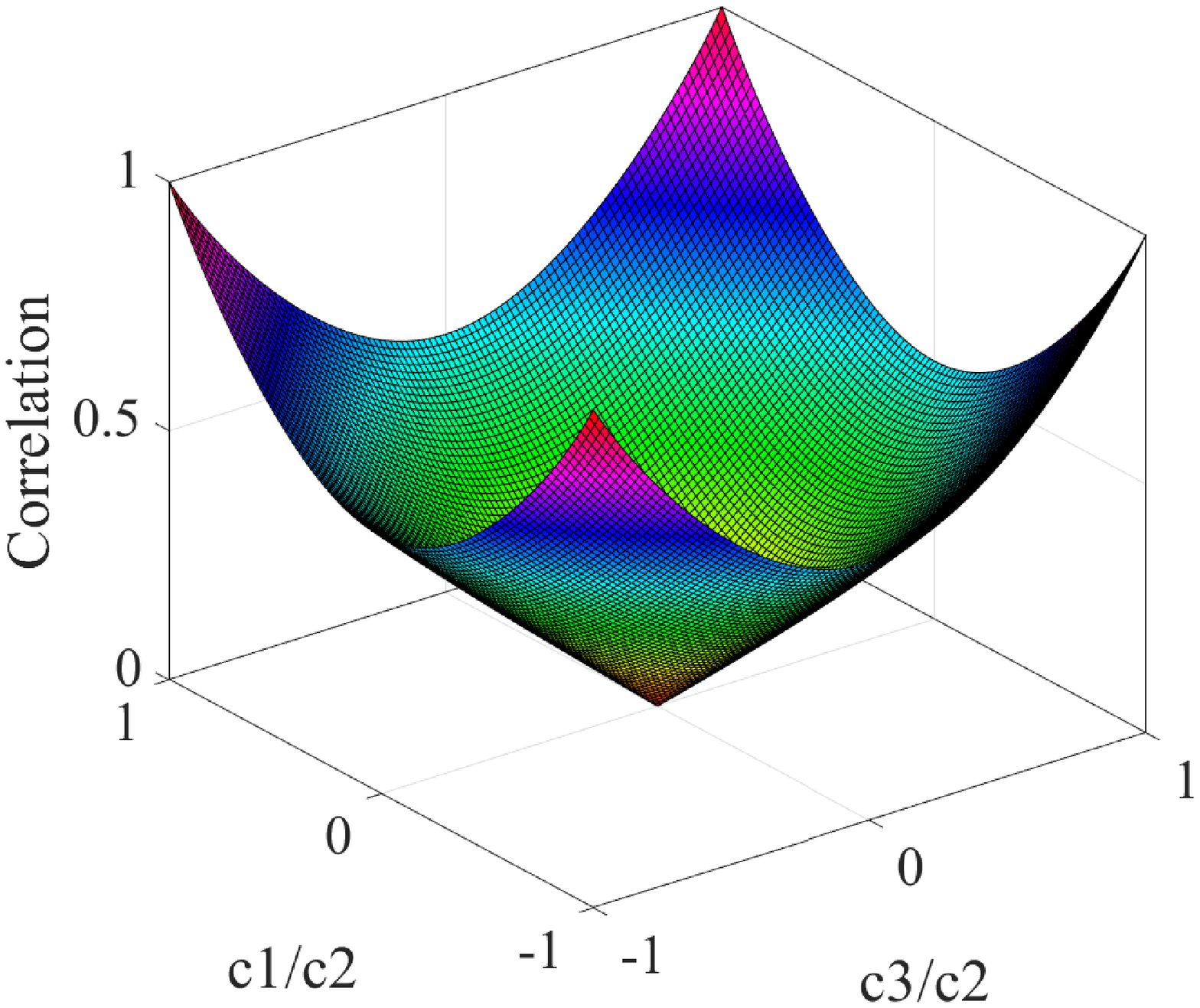}
\caption{} \label{Fig1a}
\end{figure}

\newpage
Fig. 3. The Non-local extractable works $W_{Non-loc}$ (solid line) and $W^{\{\hat{\Pi}_{i,j}\}}_{Non-loc}$ (dashed line) divided by $\Delta\varepsilon$ of Bell-diagonal state for the case $c_{1}=\frac{1}{2}, c_{2}=-\frac{1}{2}$  and  $c_3=\mathrm{sin}(\theta)$.
\begin{figure}
\centering
\includegraphics[width=300 pt]{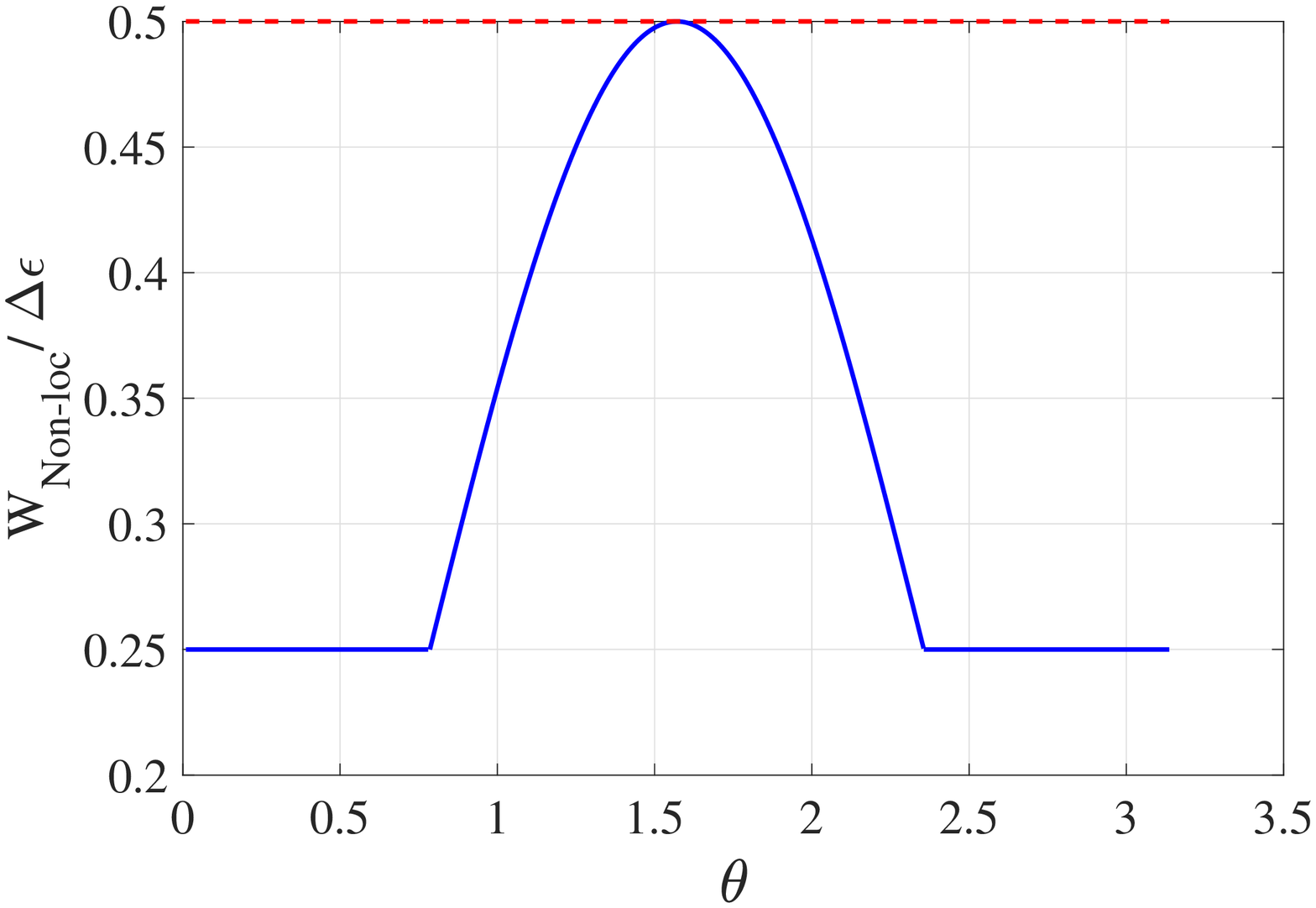}
\caption{} \label{Fig1a}
\end{figure}

\newpage
Fig. 4. The quantum correlation of Bell-diagonal state for the case $c_{1}=\frac{1}{2}, c_{2}=-\frac{1}{2}$  and  $c_{3}=\mathrm{sin}(\theta)$.
\begin{figure}
\centering
\includegraphics[width=300 pt]{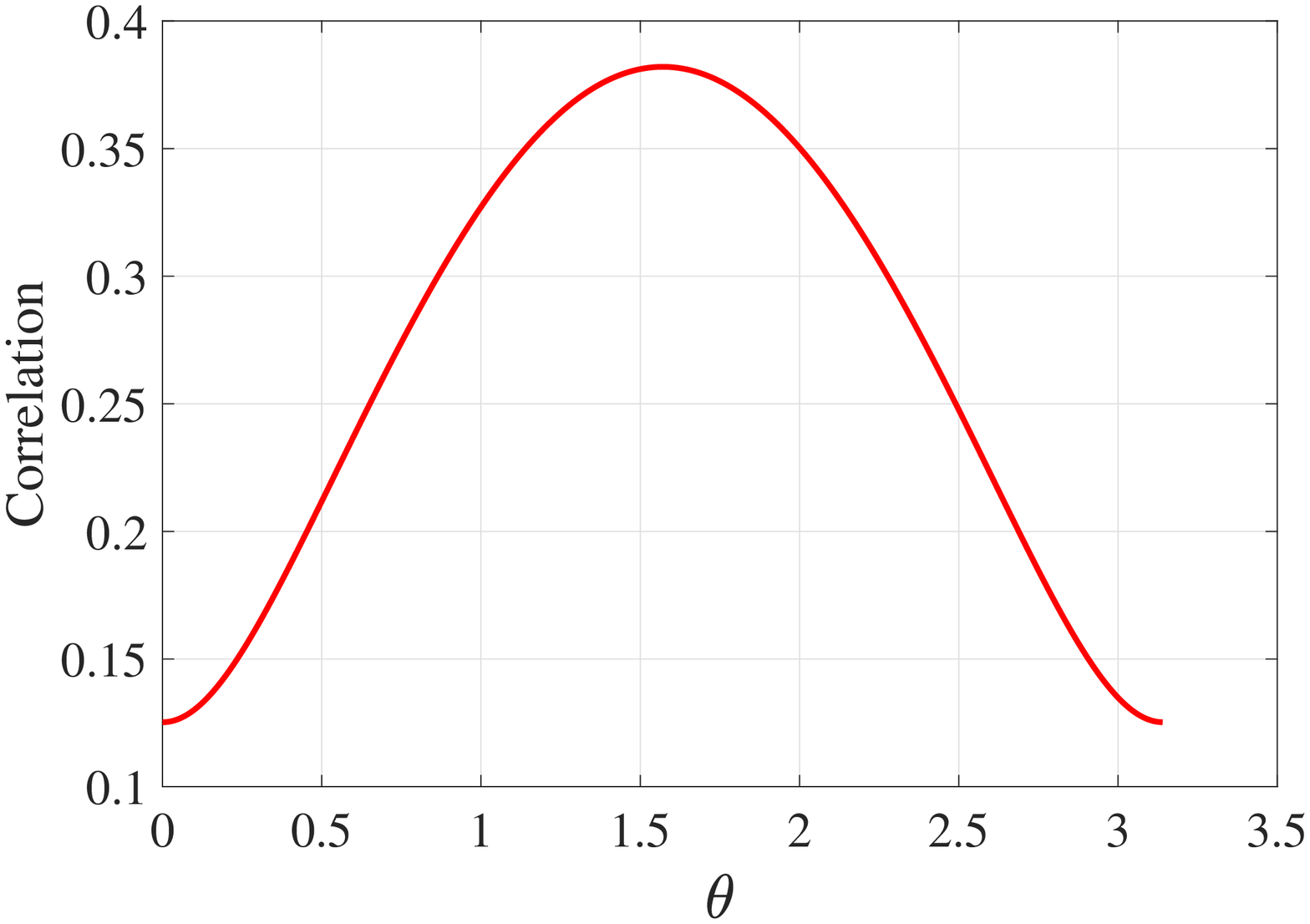}
\caption{} \label{Fig1a}
\end{figure}

\end{document}